# Roadmap for Cybersecurity in Autonomous Vehicles

**Vipin Kumar Kukkala, Sooryaa Vignesh Thiruloga, and Sudeep Pasricha**
Colorado State University

*Abstract*— **Autonomous vehicles are on the horizon and will be transforming transportation safety and comfort. These vehicles will be connected to various external systems and utilize advanced embedded systems to perceive their environment and make intelligent decisions. However, this increased connectivity makes these vehicles vulnerable to various cyber-attacks that can have catastrophic effects. Attacks on automotive systems are already on the rise in today's vehicles and are expected to become more commonplace in future autonomous vehicles. Thus, there is a need to strengthen cybersecurity in future autonomous vehicles. In this article, we discuss major automotive cyber-attacks over the past decade and present state-of-the-art solutions that leverage artificial intelligence (AI). We propose a roadmap towards building secure autonomous vehicles and highlight key open challenges that need to be addressed.**

## I. INTRODUCTION

THE aggressive attempts of automakers to make vehicles fully autonomous have resulted in increased software and hardware complexity across automotive subsystems. Many state-of-the-art automotive subsystems for collision avoidance, lane keep assist, pedestrian and traffic sign detection, etc., demand powerful embedded systems, typically referred to as electronic control units (ECUs), to be integrated into the vehicles. To meet the needs across various subsystems, a diverse set of ECUs consisting of different compute and memory capacities are used in today's vehicles. The ECUs are distributed across the vehicle and communicate using an in-vehicle network. Several in-vehicle network protocols are used in modern vehicles to meet the data rate, timing, and reliability requirements of automotive subsystems. Some of the most commonly used in-vehicle network protocols include controller area network (CAN), local interconnect network (LIN), FlexRay, and Ethernet. Both ECUs and in-vehicle networks are becoming more complex to satisfy emerging autonomy needs.

Additionally, a variety of automotive subsystems rely heavily on the data from external systems as shown in Fig. 1, which makes modern vehicles highly vulnerable to various security attacks. In the past decade (2010 onwards), nearly 79.6% of all automotive attacks have been remote attacks, which do not require the attacker to be within the vicinity of the vehicle [1]. A variety of attack vectors have been used including WiFi, telematics, Bluetooth, keyless entry systems, and mobile applications. We discuss many of these attacks in this article, as well as techniques that have been proposed to protect the vehicles from cyber-attacks. However, due to the overall increase of the automotive system complexity (heterogeneous ECUs, network architectures/protocols, and applications), detecting cyber-attacks is not easy, which poses a major challenge for emerging connected and autonomous vehicles (CAVs). There is a critical need for a monitoring solution that can serve as an intrusion detection system (IDS) to detect cyber-attacks in vehicles. Traditionally, such IDSs in computing systems have relied on using firewalls, or rule-based systems to detect cyber-attacks. These simple systems cannot detect highly complex modern automotive attacks.

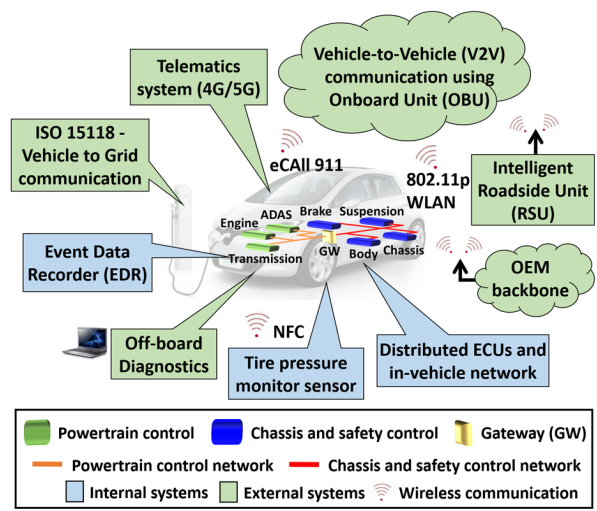

Fig. 1. Illustration of an in-vehicle network connecting different ECUs using isolated networks that are connected to a gateway (GW), and external connectivity of modern vehicles with various systems in the environment.

Another interesting trend in modern vehicles is the widespread adoption of AI techniques for advanced driver assistance subsystems (ADAS), where environmental perception is required [2]. Such AI techniques can also be deployed in powerful automotive ECUs to monitor and detect cyber-attacks. AI-based solutions are well known to be highly efficient in learning the complex patterns that exist in high dimensional time-series vehicular network data. They can observe for anomalous patterns on in-vehicle networks that connect all in-vehicle and external subsystems, to detect cyber-attacks. With fully autonomous vehicles supporting increased connectivity to external subsystems on the horizon, having an efficient IDS that can detect a variety of cyber-attacks using AI techniques is crucial and an urgent requirement.

In this article, we discuss the timeline of major automotive cyber-attacks and then present the state-of-the-art efforts that utilize advanced AI and deep-learning models to detect and address cyber-attacks in vehicles. We also present a roadmap for achieving cybersecurity goals in autonomous vehicles and present insights on the future challenges in this area. This article is motivated by the increasing frequency of attacks on automotive platforms in recent years. The emphasis on contemporary issues with cybersecurity in autonomous vehicles provides valuable insights for a general technical audience, as well as cyber-physical architects, security researchers, automakers, and Tier-N OEMs to understand the challenges in this area and positively impact the design of future security-aware automotive products.

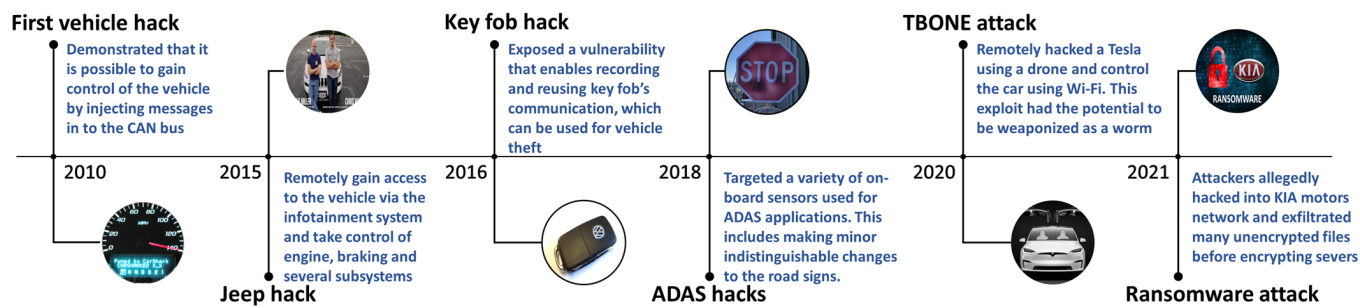

Fig. 2. Timeline of major automotive cyber-attacks.

## II. History of Automotive Cyber-Attacks

Modern automotive systems are facing an inflection point as they become highly vulnerable to various security, privacy, and safety risks. Several automotive attacks have been observed in the past that ranged from targeting a single stationary vehicle to a fleet of vehicles on the road. Here we present a timeline of the major automotive attacks (Fig. 2) and discuss their impacts.

The researchers at the University of California at San Diego and the University of Washington demonstrated one of the first vehicle hacks in 2010 [3]. They exploited the onboard telematics system and reverse-engineered the system, and were able to gain full control of the vehicle. The researchers however worked with the manufacturer and did not disclose the full details of the vulnerability. Needless to say, this hack opened up a Pandora's box. Several other works followed this approach and tried to reverse engineer the ECUs in the vehicles in the following years. However, all of these attacks required the attacker to be physically present inside the target vehicle, which resulted in dismissing as an unlikely situation and these hacks not gaining much traction in the media.

This changed in 2015 when the first major remote attack was demonstrated on an unaltered 2014 Jeep Cherokee [4] by two security researchers. The researchers identified a software bug in the vehicle's infotainment system that would allow them to connect to the vehicle remotely over the 4G LTE, and send CAN messages to the ECUs in the vehicle. They demonstrated a wide range of attacks ranging from remotely controlling simple functionality such as the vehicle radio, A/C, and windshield wipers to more critical functionalities such as controlling brakes, transmission, and even killing the engine while the vehicle was on a freeway. The attackers were able to launch these attacks on a remote vehicle from their home. This hack created a huge media outburst and the manufacturer had to issue a patch to fix the bug. The same researchers in 2016 exposed another bug that let them remotely control the acceleration, steering wheel, and cruise control systems. Similar attacks came into light in 2016, where an attacker was able to remotely control a Nissan Leaf in England from Australia. These attacks changed the landscape of how automotive cyber-attacks were carried out and highlighted the urgency to address cybersecurity in vehicles.

Starting around 2016, a new type of attack emerged that focused on hacking the keyless entry system in vehicles. The goal of these attacks was to steal the vehicle rather than remotely control it. The researchers at the University of Birmingham showed how they were able to recover the cryptographic algorithms and keys from the ECUs and clone the VW group remote control by eavesdropping on a single signal sent by the original remote [5]. Several other attacks have emerged that targeted mobile applications that were used for remote start and immobilizer systems in vehicles. Some of the recent attacks in this class include cloning the Tesla Model S key fob in 2018 [6] and the Tesla Model X key fob in 2019 [7]. Researchers were able to clone key fobs by capturing the Bluetooth communication between the key fob and the body control module (BCM) and were able to use a bootleg BCM to replay it and steal the vehicle in under 90 seconds.

A different class of attacks has gained popularity since 2018, mainly targeting the ADAS systems and the onboard sensors used for perception. In [8], researchers generated various robust visual adversarial perturbations to a stop sign that resulted in it being misidentified as a 45 mph speed limit sign. A few years before this, researchers were able to blind a Mobileye C2-270 camera and demonstrate jamming, spoofing, and relay attacks on an Ibeo LUX3 LiDAR sensor [9]. More recent attacks include tricking lane change system of a Tesla Model S with bright stickers on the road by Tencent Keen security lab in 2019 [10] and object removal attacks on LiDAR sensors in 2021 [11]. Another recent attack that made the headlines was the T-BONE attack [12] where researchers were able to gain remote code execution (RCE) over WiFi on the infotainment system in a Tesla Model 3 using a drone. They were able to remotely open doors and trunk, change seat positions, steering, and acceleration modes. However, this exploit does not provide driving control of the vehicle. The researchers also highlighted that adding a privilege escalation exploit such as CVE-2021-3347 to T-BONE would weaponize this exploit and turn it into a worm. This would allow them to load new WiFi firmware and exploit other Tesla cars in the victim car's proximity. More recently, in the beginning of 2021, an online hacking group by the name DoppelPaymer claimed to conduct a ransomware attack on KIA motors America and have stolen unencrypted confidential data [13].

Thus, cyber-attacks are becoming increasingly prevalent in modern vehicles. A comprehensive taxonomy of vehicular security attacks with details related to attacker types, attack tools and actions, and attack objectives are discussed in [31].

## III. AI To The Rescue for Intrusion Detection in Vehicles

To protect vehicles from devastating cyber-attacks, it is crucial to address cybersecurity in the automotive domain. Several lightweight authentication mechanisms such as [25] and [33], and versatile automotive security frameworks such as [26] try to improve security in resource-constrained automotive ECUs. However, these techniques mainly focus on authenticating the ECUs during the initial setup phase and cannot detect the presence of an attacker in the later stages or when an already authenticated ECU gets compromised. Thus, there is a need for an intrusion detection system (IDS) that continuously monitors the vehicular network and detects any cyber-attacks. Moreover, an IDS is extremely crucial and often the last line of defense when the attacker breaks through defense mechanisms. Traditional IDS solutions relied on firewalls or rule-based (non AI-based) systems to detect cyber-attacks. However, they are not effective against detecting sophisticated automotive attacks as they fail to capture the complex dependencies in the time-series vehicular network data. With the availability of large troves of data in vehicles from communication between ECUs and external systems, and the increased computing capabilities of ECUs, AI-based solutions can be leveraged to parse high dimensional vehicular network data. Here we discuss some recent work on AI-based IDSs that encompass monitoring and attack detection at an in-vehicle network level and vehicular ad hoc network (VANET) level.

### A. IDSs for In-Vehicle Network Security

A generative adversarial neural network (GAN) based IDS was introduced in [14] that tries to learn the patterns of the message identifier (ID) in CAN data, by transforming the ID data into "CAN images", to detect attacks. This approach employs two discriminator models to detect known and unknown attacks, respectively. The former is trained using normal and abnormal CAN images from actual vehicle data, and the latter is trained simultaneously using the adversarial process (a mix of adversarial data from the generator and normal CAN images). The second discriminator learns to detect fake CAN image that looks similar to real CAN images. This approach showed high detection accuracy against distributed denial of service (DDoS), and message injection attacks.

A gated recurrent unit (GRU) based recurrent autoencoder IDS is presented in [18]. During training, this technique learns the normal operating behavior of the system by reconstructing the input attack-free CAN data. During inference at runtime, this technique observes the deviation from the learned normal system behavior to detect cyber-attacks. This approach was implemented on real-world automotive hardware and tested under several attack scenarios. In [16], an unsupervised LSTM-based encoder-decoder was proposed that integrates a novel self-attention mechanism to learn the characteristics of normal data traversing the in-vehicle network. The attention mechanism enhanced the ability of the model to focus on the important hidden state information from the past. A one-class support vector machine (OCSVM) based classifier was trained and deployed together with the self-attention model at runtime to detect cyber-attacks. This approach demonstrated superior performance in accuracy, F1 score, false-positive rate (FPR), and receiver operating characteristics (ROC) curve-area under the curve (AUC) compared to various state-of-the-art statistical, proximity-based, and machine learning (ML) based IDS approaches under different attacks scenarios. In [15], a weighted state graph (WSG) was constructed in the offline phase with the CAN message IDs as the vertices of the graph and the edges of the graph representing the time-varying state transitions of CAN data frames. At runtime, a sub-graph was generated for a sliding window of CAN IDs, and an anomaly was triggered if the generated sub-graph was not within the range of the WSG created during training. This technique focuses on detecting a wide range of attacks and minimizing FPR. A temporal convolutional neural attention (TCNA) network-based lightweight IDS was presented in [17], to learn very long-term dependencies between messages in an in-vehicle network. This was used to effectively learn the normal system behavior of the vehicle at design time. A decision tree (DT) based classifier was used to learn the model deviations that correspond to the normal vehicle operation at design time. At runtime, the trained TCNA and DT models were used to observe for deviations using the deviation score metric to detect cyber-attacks. This technique outperformed all the comparison works in the Matthews correlation coefficient (MCC) metric.

### B. IDSs for VANET Security

Several other efforts have focused on AI-based IDS in VANET environments, particularly during vehicle-to-vehicle (V2V), and vehicle-to-infrastructure (V2I) communication.

An AI-based IDS framework was introduced in [21] to detect and mitigate anomalous vehicle responses of cooperative adaptive cruise control (CACC) modules in a V2V network. An ML-based predictor was trained to predict the acceleration of the preceding vehicle, and an anomaly alert was raised when the observed acceleration exceeded the prediction by a certain threshold. The framework then assessed the threshold deviation level and would override the CACC when anomalies were detected. On-road driving data such as the preceding vehicle acceleration, inter-vehicle space, vehicle velocity was periodically collected to train the ML model. A data-driven IDS to detect attacks occurring in roadside units (RSUs) communicating with a vehicle was presented in [19]. It used a convolution neural network (CNN) based architecture to extract the link load features of the RSU to detect attacks. This technique demonstrated high accuracy compared to neural network, support vector machine, and principal component analysis models. However, this technique was evaluated only a limited set of attacks. A hybrid IDS framework was proposed in [20] that performed a three-phase data traffic analysis, reduction, and classification on the received cloud service requests to detect intrusions. A combination of a deep belief neural network (DBN) and a DT classifier was presented to perform data dimensionality reduction and attack classification. This technique achieved high detection accuracy and lower FPR when evaluated under DoS, probe, remote to user (R2L), and user to root (U2R) attacks. However, this approach incurs large detection accuracy. An LSTM based autoencoder model was proposed in [22] to detect intrusion in V2X networks. The LSTM model tries to learn the temporal representation of the compressed network traffic data, and any deviation from the learned representation is classified as an attack. The authors claim their technique can detect intrusions in both in-vehicle and external networks with high accuracy.

TABLE I
SUMMARY OF AI-BASED IDS SOLUTIONS DISCUSSED IN THIS ARTICLE

| Level | Work | Technique |
|---|---|---|
| In-vehicle network | GIDS [14] | GAN based IDS using CAN images |
| | INDRA [18] | GRU based recurrent autoencoder; static threshold based attack detection |
| | LATTE [16] | LSTM based encoder-decoder with self-attention; OCSVM based attack detection |
| | TENET [17] | Temporal CNN with neural attention; DT based classifier for attack detection |
| VANET | RACCON [21] | Explores five different machine learning techniques; static threshold to detect V2V attacks |
| | DD-IDS [19] | CNNs to detect attacks aimed at RSUs |
| | CS-IDS [20] | DBN for data reduction; DT classifier to detect attacks in received cloud service requests |
| | AED-ITS [22] | LSTM based autoencoder; static threshold to detect V2V and V2I attacks |

A summary of both in-vehicle network and VANET level IDSs discussed in this article are summarized in Table I. Moreover, the techniques in [16], [17], and [18] have explored the feasibility of AI-based IDS on a real-world automotive ECU. They were able to detect various cyber-attacks with minimal computational, memory, and power overhead.

## IV. CYBERSECURITY ROADMAP FOR AUTONOMOUS VEHICLES

Despite the many promising state-of-the-art AI-based IDS techniques that show compelling results, several issues still need to be addressed to make future autonomous vehicles truly secure. Here we discuss various key elements in roadmap (shown in Table II) to achieve autonomous automotive cybersecurity goals in the near future. Note that the order in which the key elements in the roadmap are presented does not represent a recommendation for solving them sequentially in that order. These key elements can be addressed concurrently.

### A. Cybersecurity Aware Design Practices

Traditional automotive systems are designed by having safety and reliability requirements as the main focus, and security is added as an additional layer on the top. This severely limits the scope of security goals that can be realized, which exposes the automotive systems to a variety of attacks. On the other hand, in a cybersecurity-aware design process, the security requirements are considered from the beginning of the development cycle. This integration of security as one of the fundamental requirements in the design process helps in achieving a diverse set of security goals, making automotive systems robust and better protected against cyber-attacks.

Additionally, advanced security strategies such as the defense-in-depth or castle strategy [27], where a series of layered security mechanisms are built to provide confidentiality and integrity goals, need to be adapted to further enhance security. These approaches introduce redundancy in the defense mechanisms that can be highly effective against complex attacks.

The modern automotive system design process also needs to employ proactive security strategies such as a zero-trust security model. This model relies on never trust, always verify principle, where no participant or transaction is trusted by default, even if they were verified earlier. Zero trust practices advocate for mutual authentication and verifying their identity and integrity. This is a prominent strategy in cloud-based systems, data centers, and corporate IT, that needs to be customized for time-critical resource-constrained automotive systems. Lastly, designing lightweight security protocols will be crucial, especially with the adoption of 5G and newer networks in VANETs that will require security-specific processing to finish within microsecond latencies to avoid missing deadlines.

TABLE II
KEY ELEMENTS IN CYBERSECURITY ROADMAP FOR AUTONOMOUS VEHICLES

| Roadmap Elements | Components |
|---|---|
| Cybersecurity-aware design practices | • Security requirements<br>• Multi-layered security<br>• Zero trust security |
| Secure hardware and software stack | • HSMs and TPMs<br>• SDL for automotive software<br>• SOTA and FOTA updates |
| New security and AI standards and regulations | • ISO/SAE 21434<br>• UNECE WP. 29<br>• AI regulations |
| Advanced threat intelligence | • Vulnerability assessment<br>• Penetration testing<br>• Auto ISAC |
| Open challenges | • Data protection and privacy<br>• Tamper-proof AI<br>• Securing automotive IC supply chain<br>• Adopting emerging technologies |

### B. Secure Hardware and Software Stack

With level-3 and level-4 autonomous vehicles involving high and full automation (as defined in SAE J3016 [34]) expected to hit the roads soon, automotive hardware and software is experiencing a paradigm shift. Modern ECUs handle various safety-critical applications such as autonomous lane change, collision avoidance, airbag deployment, etc., which makes them an attractive target for malicious attackers. One of the popular approaches to protect ECUs from being attacked is by implementing security mechanisms to authenticate and verify the identity of the components. However, this incurs high overhead on the ECUs and can jeopardize real-time operations of the vehicle. To address this, future autonomous vehicles must utilize cryptographic accelerators called hardware security modules (HSMs) to offload security tasks. Moreover, future vehicles need to employ trusted platform modules (TPMs), which provide a hardware root of trust to enable secure computing and secure key storage. A comprehensive discussion of hardware security primitives for vehicles including HSMs and TPMs is discussed in [32]. Moreover, recent works such as [33] that explored the use of physically unclonable functions (PUFs), and [26] that used meta-heuristics based key-management can help in achieving lightweight secure ECU authentication.

Various other security features will also need to be supported within ECUs, such as core isolation (which protects safety-critical processes by isolating them in memory), memory integrity (to ensure malicious code is not injected in the memory corresponding to critical processes), and address space layout randomization (to randomize memory locations where the applications are loaded and protect against buffer overflow attacks). Moreover, automotive application development should adopt security development lifecycle (SDL) [28] practices that involve including the security artifacts in the software development lifecycle. Automakers need to adopt the over the air (OTA) updates as a standard practice to provide patches to the software (SOTA) and firmware (FOTA) vulnerabilities and need to include the required hardware to enable this. Currently, SOTA updates are employed by multiple automakers but FOTA updates remain uncommon. Lastly, adopting secure multi-party computation (MPC) eliminates the need for sharing data with a trusted third-party and allows collaborators to work securely on encrypted data. In 2021, Bosch launched an open-source project called Carbyne stack [29] that enables secure data processing using MPC.

### C. New Security and AI Standards and Regulations

In August 2021, ISO and SAE jointly published the ISO/SAE 21434 standard for Road Vehicles - Cybersecurity Engineering. It specifies various engineering requirements for cybersecurity risk management for road vehicles including many of their components and interfaces. Another standard, SAE J3061 provides guidance and establishes high-level principles related to security of cyber-physical systems. There is a need for the development of similar standards across all vehicle subsystems, and to encourage OEMs and Tier-N suppliers to adopt them. The lack of current cybersecurity-centric regulatory mandates allows automakers today to have less stringent security requirements resulting in subpar security in vehicles. In 2020, the world forum for harmonization of vehicle regulations (WP. 29) of the United Nations Economic Commission for Europe (UNECE) adopted two new regulations that mandate all automakers in the 54 contracting nations to adopt cybersecurity management systems (CSMS) and software update management systems (SUMS) by July 2022.

Moreover, with the increasing adoption of AI algorithms in modern vehicles, there is a growing need for regulations around AI development. In April 2021, the European Commission introduced the first ever legal regulation on AI to mitigate the hazards of "high risk" AI applications. Even though this does not specifically focus on autonomous vehicles, this is a move in the right direction. There is also a need for more legal regulations around crashes and other accidents caused by driving decisions made by AI algorithms in autonomous vehicles. The need for this was highlighted in 2018 when a ridesharing company's autonomous vehicle testing resulted in a fatal crash leading to the death of a pedestrian. The company did not face any criminal charges but halted testing of the autonomous vehicles.

### D. Advanced Threat Intelligence

Threat intelligence refers to the act of collecting, processing, and analyzing data to understand existing and future threats. It is crucial to continuously monitor vehicles for any cyber-threats, to design effective remedial actions. Vehicles today go through multiple years of rigorous testing to ensure safety functionalities but do not go through any comparable testing to ensure security. Threat intelligence data needs to be collected for vehicles and used for enhanced cybersecurity testing. Conducting regular vulnerability assessments and penetration testing on vehicles will be crucial and should be made just as important (and frequent) as regular vehicle maintenance. It is also important to have a common platform to share this information between different organizations to effectively tackle cyber-attacks. To solve this, the automotive information sharing and analysis center (Auto-ISAC) was formed in 2015. It consists of 52 members ranging from OEMs to Tier-N suppliers actively sharing information about cyber-threats in vehicles. Auto-ISAC also works with the U.S. Department of homeland security (DHS) to share vulnerabilities with the federal government. Such advanced threat intelligence sharing across organizations will be crucial to ensure the security of future autonomous vehicles.

## V. OPEN CHALLENGES

Beyond the recommendations that have already been discussed, we present some key open challenges that represent promising opportunities for researchers to assist with achieving security goals in future autonomous vehicles.

### A. Data Protection and Privacy

Data theft is a rapidly growing concern in today's world and is a prevalent issue across various industries. In 2020 alone, the average cost of a single data breach was around $3.86 million [1]. This is also a concern in future autonomous vehicles as the vehicles collect and operate on large volumes of data of different types. Data thefts have varying levels of safety, security, and economic impacts depending on the type and severity of the breach. Such thefts can compromise individual user data as well as intellectual property data of vehicle OEMs. With the recent introduction of next-generation driverless ridesharing services in places such as Phoenix and San Francisco, the stakes for user data privacy now is higher than ever. For instance, attackers can use stolen user information to launch more effective socially engineered attacks. The issues of security, trust, and privacy in autonomous vehicles are presented in detail in [24]. Techniques such as confidentiality integrity availability (CIA) and distributed immutable ephemeral (DIE) models need to be adopted in the automotive domain to ensure data protection and privacy of future autonomous vehicles.

### B. Tamper-proof AI

AI algorithms have shown superior performance in IDS and ADAS subsystems for autonomous vehicles. However, these algorithms are vulnerable to carefully crafted adversarial attacks [8]. Moreover, with the rollout of increasingly connected vehicles, we envision that Black-hole DDoS attacks (where communication between vehicles is blocked) and Sybil attacks (where a vehicle operates with multiple identities) will become increasingly common. Such attacks will result in confusing AI algorithms, potentially causing failure across vehicle subsystems. Recent model inversion attacks [30] that try to reconstruct training data from the model parameters are gaining popularity. Such attacks pose a great threat to the proprietary data of the automakers that are used to train the AI models. Moreover, with newer and scalable learning approaches for large AI algorithms such as with federated learning in datacenter environments, the need for creating new approaches for tamper-proof and adversarial attack-resilient AI algorithms becomes even more imperative. A comprehensive survey of adversarial attacks on AI/ML algorithms in CAVs, with adversarial defense approaches and open challenges is discussed in detail in [35].

### C. Securing Automotive IC Supply Chains

As different semiconductor integrated circuit (IC) components in a vehicle are manufactured in different parts of the world today, it is crucial to have a secure supply chain. Any vulnerability induced from the supply chain in any component of the vehicle will have disastrous effects on autonomous vehicles. This issue is further exacerbated with the increasing demand for the RSUs and 5G infrastructure to enable intelligent transportation systems. A comprehensive list of IC supply chain concerns and a logic obfuscation technique to overcome them is presented in [23]. Techniques such as digital watermarking, IC fingerprinting, IC metering, etc., need to be further explored to ensure a secure supply chain.

### D. Adopting Emerging Technologies

In recent years, researchers have started looking into using WiGig networks that use IEEE 802.11ad multiple gigabit wireless system (MGWS) standard at 60 GHz frequency for in-vehicle network communication. The ability to support high data rates and enable low latency applications can transform the prospect of both in-vehicle networks and future self-driving applications. The feasibility of using IEEE 802.11ad millimeter wave (mmWave) for in-vehicle communication between ECUs is demonstrated in [36], with a worst-case (experimentally-observed) throughput of around 300 Mbps. Another disruptive technology that could revolutionize future autonomous vehicles is blockchain technology. The blockchain's decentralized ledger provides accurate and simultaneous access to different types of data, such as traffic information and better vehicle tracking information for ride-sharing applications. A blockchain-based scheme to mitigate the security and privacy issues in autonomous vehicles is discussed in [37]. However, as these technologies are still in their infancy in the automotive domain, they need to be meticulously scrutinized by characterizing vulnerabilities and exploring security mechanisms to enhance security.

## VI. CONCLUSION

In this article, we presented a comprehensive timeline of the major automotive cyber-attacks and discussed their impact. We then discussed the need for efficient intrusion detection systems and presented various state-of-the-art AI-based intrusion detection techniques. We presented a roadmap for realizing secure autonomous vehicles in the future, while discussing both technical and regulatory issues that need to be addressed along with the key open challenges that currently remain unsolved. The key elements of the cybersecurity roadmap and open challenges highlight some of the crucial concerns that need to be addressed to solve the problem of cybersecurity in future autonomous vehicles. Moreover, secure hardware and software stacks in conjunction with advanced threat intelligence will be the vital fundamental components towards realizing various security goals. Ultimately, incorporating cybersecurity-aware vehicle design practices, as discussed in this article, will be of utmost importance to comprehensively achieve security goals required to safeguard future vehicles from emerging sophisticated attacks across the entire vehicle infrastructure.

### ACKNOWLEDGMENTS

This work was supported by National Science Foundation (NSF), through grant CNS-2132385.